\documentclass[sigconf]{acmart}

\usepackage{multirow}

\AtBeginDocument{%
  }

\makeatletter
\def\@ACM@copyright@check@cc{}
\makeatother

\copyrightyear{2025}
\acmYear{2025}
\setcopyright{cc}
\setcctype{by}
\acmConference[FAccT '25]{The 2025 ACM Conference on Fairness, Accountability, and Transparency}{June 23--26, 2025}{Athens, Greece}
\acmBooktitle{The 2025 ACM Conference on Fairness, Accountability, and Transparency (FAccT '25), June 23--26, 2025, Athens, Greece}\acmDOI{10.1145/3715275.3732130}
\acmISBN{979-8-4007-1482-5/2025/06}

\begin{document}
\title[Centralized Trust in Decentralized Systems]{Centralized Trust in Decentralized Systems: Unveiling Hidden Contradictions in Blockchain and Cryptocurrency}

\author{Faisal Haque Bappy}
\affiliation{%
   \institution{Syracuse University}
   \city{Syracuse}
   \state{NY}
   \country{USA}}
\email{fbappy@syr.edu}

\author{EunJeong Cheon}
\affiliation{%
   \institution{Syracuse University}
   \city{Syracuse}
   \state{NY}
   \country{USA}}
\email{echeon@syr.edu}

\author{Tariqul Islam}
\affiliation{%
   \institution{Syracuse University}
   \city{Syracuse}
   \state{NY}
   \country{USA}}
\email{mtislam@syr.edu}

\renewcommand{\shortauthors}{Bappy et al.}

\begin{abstract}
  Blockchain technology promises to democratize finance and promote social equity through decentralization, but questions remain about whether current implementations advance or hinder these goals. Through a mixed-methods study combining semi-structured interviews with 13 diverse blockchain stakeholders and analysis of over 3,000 cryptocurrency discussions on Reddit, we examine how trust manifests in cryptocurrency ecosystems despite their decentralized architecture. Our findings uncover that users actively seek out and create centralized trust anchors, such as established exchanges, prominent community figures, and recognized development teams, contradicting blockchain's fundamental promise of trustless interactions. We identify how this contradiction arises from users' mental need for accountability and their reluctance to shoulder the full responsibility of self-custody. The study also reveals how these centralized trust patterns disproportionately impact different user groups, with newer and less technical users showing stronger preferences for centralized intermediaries. This work contributes to our understanding of the inherent tensions between theoretical decentralization and practical implementation in cryptocurrency systems, highlighting the persistent role of centralized trust in supposedly trustless environments.
\end{abstract}

\settopmatter{printacmref=true}




\maketitle

\section{Introduction}
Blockchain technology and cryptocurrencies are often celebrated as transformative innovations that promise decentralization, transparency, and user empowerment \cite{swan2015blockchain, zheng2018overview}. At their core, these systems aim to eliminate intermediaries, democratize access to financial tools, and grant users unprecedented control over their assets. However, the real-world adoption of blockchain technologies has exposed significant challenges that undermine these ideals. Users frequently encounter steep learning curves, inaccessible interfaces, and security vulnerabilities, which lead them to rely on centralized platforms for convenience and support \cite{de2018cryptocurrencies, prewett2020blockchain}. This reliance on centralized entities contradicts the decentralized ethos of blockchain, raising important questions about the fairness, accountability, and transparency of these systems in practice \cite{xiao2024centralized21, dupuis2021blockchain}.

Recent trends in the cryptocurrency market highlight a growing tension between blockchain's revolutionary potential and its practical realities \cite{web3devs2024}. The meteoric rise and subsequent collapse of platforms such as FTX have exposed systemic vulnerabilities, leading to billions of dollars in losses and shaking user confidence in the ecosystem \cite{ftx, conlon2023collapse}. Similarly, high-profile scams, including fraudulent initial coin offerings (ICOs) and rug-pulls, continue to exploit the absence of standardized regulations and oversight mechanisms in the cryptocurrency space—leaving investors vulnerable to manipulation and fraud—further eroding trust in the promise of decentralization \cite{schilling2019decentralized, scharfman2023cryptocurrency}. These events, coupled with the collapse of algorithmic stablecoins like TerraUSD \cite{krause2025algorithmic}, underscore the fragility of blockchain systems and the persistent risks to both novice and experienced users.

Despite these setbacks, blockchain technology continues to evolve, with emerging trends such as decentralized finance (DeFi), non-fungible tokens (NFTs), and central bank digital currencies (CBDCs) gaining traction \cite{auer2023rise}. However, recent studies \cite{jain2021we, huang2021rich} suggest that blockchain’s egalitarian promise remains elusive. Mining power and cryptocurrency holdings are increasingly concentrated among a small number of entities, such as large mining pools, institutional investors, and exchanges, which control significant portions of the network’s resources. This concentration creates opaque structures of centralized power that may be less accountable than traditional financial institutions \cite{de2018cryptocurrencies, gao2016two13}. Additionally, the high technical barriers and financial costs associated with meaningful participation in blockchain systems—such as the need for expensive hardware for mining or staking, and the complexity of navigating blockchain protocols—often exclude marginalized communities \cite{nelson2024bitcoin}. This exclusion exacerbates existing socioeconomic disparities, limiting access to the benefits of blockchain technology and reinforcing systemic inequalities \cite{mai2020user18, voskobojnikov2021u12}.

The disconnect between blockchain’s ideals and user behaviors is further compounded by widespread misconceptions about its functionality and risks. Even experienced users struggle with concepts such as transaction irreversibility and secure private key management, leading to costly errors and vulnerabilities \cite{gao2016two13, voskobojnikov2021u12, kristiansen2007rules}. Moreover, the centralization-decentralization paradox—where users espouse a preference for decentralization yet gravitate toward centralized platforms—reveals a critical misalignment between user needs and system design \cite{knittel2019true16}. This paradox not only undermines the foundational principles of blockchain but also raises fundamental questions about the accountability of decentralized systems in addressing user concerns.

This study investigates these dynamics through the lens of fairness, accountability, and transparency, focusing on the interplay between user behavior, system design, and the principles of decentralization. By examining the centralization-decentralization paradox, the challenges of trust formation, and the social and technical barriers to adoption, this work aims to uncover how blockchain technology can better align with user needs and expectations. To address these challenges, we employ a mixed-methods approach that combines semi-structured interviews with diverse blockchain stakeholders and a systematic content analysis of cryptocurrency discussions on Reddit. This dual methodology enables us to capture both nuanced individual experiences and broader community perspectives, providing a comprehensive understanding of the ecosystem. 

Our findings highlight persistent barriers to blockchain adoption, including inaccessible interfaces, insufficient education, and a lack of transparent metrics for evaluating blockchain projects. These challenges impede adoption, deepen inequities, and undermine blockchain's goals of fairness and accountability. This paper explores these issues, offering insights to design systems that align decentralization ideals with user needs. The following are the key contributions of this paper. 

\begin{itemize}
    \item \textbf{Illuminating User Barriers to Adoption:} Through interviews and Reddit comment analysis, we uncover critical obstacles that prevent users from fully engaging with decentralized systems. These include technical complexity, inaccessible interfaces, and the lack of reliable educational resources. By addressing these barriers, the paper provides actionable insights for improving the accessibility of blockchain technology.

    \item \textbf{Examining the Trust Paradox in Blockchain:} We explore how users navigate the tension between decentralization and their desire for accountability. Our findings reveal that while blockchain is designed to operate as a trustless system, users often prefer centralized services that offer customer support and security assurance. This insight challenges the assumption that decentralization alone is sufficient to foster trust and highlights the need for hybrid trust models that balance decentralized principles with centralized accountability.
 
    \item \textbf{Advancing Transparency in Blockchain Evaluation:} We identify the absence of standardized, transparent metrics for assessing the legitimacy and reliability of blockchain projects as a significant barrier to informed decision-making. The paper calls for the development of accessible evaluation frameworks that enable users, researchers, and developers to assess the sustainability of a project, user adoption, and technical reliability, thereby fostering accountability and reducing market volatility.

    \item \textbf{Proposing User-Centered Design Strategies:} The findings emphasize the importance of designing blockchain systems that prioritize user needs while maintaining core principles of decentralization. This includes simplifying authentication processes, enhancing user interfaces, and providing better support for secure key management. By aligning technical capabilities with practical user requirements, the paper outlines pathways for making blockchain technology more equitable and inclusive.
\end{itemize}

\section{The Centralization vs Decentralization Paradox in Cryptocurrencies}
The concept of decentralization in cryptocurrency originated with Bitcoin's groundbreaking white paper by Satoshi Nakamoto in 2008, which proposed a revolutionary solution to the long-standing problem of digital trust \cite{nakamoto2008bitcoin}. At its core, Bitcoin introduced a decentralized ledger system---the blockchain---that fundamentally reimagined how financial transactions could be verified without a central authority. The blockchain's distributed consensus mechanism ensures security through a network of independent nodes, where no single entity controls the entire system. Each transaction is verified by multiple participants, creating a transparent and tamper-resistant record that eliminates the need for traditional intermediaries like banks \cite{antonopoulos2017mastering}.

Initially, Bitcoin's decentralization model was implemented as a public blockchain, allowing anyone to participate in the network, validate transactions, and maintain the ledger. This open model contrasted sharply with traditional financial systems, where centralized institutions like banks controlled and validated all transactions. The public blockchain's transparency and security were achieved through complex cryptographic techniques and a consensus mechanism called Proof of Work, where network participants (miners) solve complex mathematical problems to validate transactions and create new blocks \cite{swan2015blockchain}. However, the cryptocurrency ecosystem quickly evolved to include private blockchain systems, particularly for enterprise and institutional applications. These private blockchains maintain the core principles of distributed ledger technology but restrict participation to pre-approved entities. Companies like IBM \cite{IBMHyperledger} and R3 \cite{brown2016corda} have developed private blockchain solutions for industries ranging from supply chain management to financial services, offering enhanced privacy and controlled access while retaining some decentralization principles \cite{zheng2018overview}.

The emergence of cryptocurrency exchanges represents a critical inflection point in the centralization-decentralization debate. Despite marketing themselves as champions of decentralized finance, most major exchanges operate with profound centralization. Platforms like Coinbase, Binance, and Kraken present themselves as guardians of blockchain's decentralized ethos while maintaining complete control over user funds, transaction processes, and platform governance \cite{butler2018cryptocurrency}. These exchanges employ sophisticated policy mechanisms that create an illusion of decentralization while maintaining strict centralized control. For instance, they often: 1) use blockchain terminology extensively to create a perception of decentralization \cite{werbach2018blockchain}, 2) implement complex user agreements that centralize platform control (e.g., terms of service that allow exchanges to freeze accounts or restrict withdrawals at their discretion) \cite{chohan2017decentralized}, 3) maintain sole custody of user private keys \cite{preukschat2018blockchain}, 4) create opaque governance structures (e.g., Binance's lack of transparency in its decision-making processes or Coinbase's limited user representation in policy changes) with limited user input \cite{de2018cryptocurrencies}, and 5) implement centralized listing processes—where exchanges unilaterally decide which cryptocurrencies to list, often influenced by factors such as listing fees, market demand, or partnerships—contrary to the decentralized ethos \cite{schilling2019decentralized}. The regulatory system has further complicated this dynamic. While initial cryptocurrency platforms emerged as largely unregulated spaces, government interventions have increasingly pushed exchanges toward more centralized models. Regulations like the United States Bank Secrecy Act and Know Your Customer (KYC) requirements have fundamentally transformed cryptocurrency exchanges, forcing them to implement centralized identity verification and transaction monitoring processes \cite{foley2019sex}. The tension between centralization and decentralization in cryptocurrency becomes particularly evident in ownership and control structures \cite{kusmierz2022centralized}.

This centralization trend contradicts the original vision of cryptocurrencies as democratized financial systems. The promise of individual financial sovereignty has been gradually eroded by institutional control, regulatory pressures, and the practical challenges of maintaining truly decentralized systems at scale \cite{de2018cryptocurrencies}. Emerging solutions like decentralized exchanges (DEXs) and decentralized finance (DeFi) platforms attempt to address these challenges. However, they still face significant hurdles in user adoption, interface complexity, and regulatory compliance \cite{schilling2019decentralized}. Furthermore, despite users' ideological support for decentralization, many continue to gravitate toward centralized platforms due to convenience, trust, and usability concerns, as observed in NFT marketplaces and other blockchain-based systems \cite{xiao2024centralized21, yu2024don10}. Similarly, in the context of Central Bank Digital Currencies (CBDCs), security often outweighs privacy concerns, illustrating the broader tendency for centralized solutions in certain institutional contexts \cite{abramova2023can19}.

\section{Related Work}
Prior research on blockchain and cryptocurrency adoption has identified a multifaceted aspect in which users' mental models, trust mechanisms, and security behaviors often contradict the fundamental principles of decentralized technologies. At a conceptual level of blockchain and cryptocurrency adoption, the literature can be classified into three overarching themes.

\subsection{User Understanding and Misconceptions}
Studies \cite{gao2016two13, saldivar2023blockchain1, lyke2023exploring8} have shown that users often struggle to develop accurate mental models of blockchain technology, which relies on a decentralized, transparent, and secure ledger to record and verify transactions without a central authority. Through interviews with Bitcoin users and non-users, Gao et al. \cite{gao2016two13} found that even active users were not well-versed in how the protocol functions and held misconceptions about transaction privacy, specifically around anonymity levels and data visibility. This knowledge gap extends beyond cryptocurrencies to the broader blockchain ecosystem, where Saldivar et al. \cite{saldivar2023blockchain1} examined the broader blockchain ecosystem and documented specific challenges users face with blockchain-based applications, such as difficulty managing wallets, verifying transactions, and interacting with smart contracts.

These misconceptions can have serious consequences. Voskobojnikov et al. \cite{voskobojnikov2021u12} analyzed 45,821 app reviews of mobile cryptocurrency wallets and found that users often mistakenly believe transactions are reversible. While their large-scale analysis offers valuable insights into user behavior, it may overlook deeper motivations like mental models and decision-making processes. Mai et al. \cite{mai2020user18} showed that current cryptocurrency tools fail to protect users from these misconceptions, revealing vulnerabilities in transaction security. However, their study doesn't fully address how social factors, such as trust and peer networks, influence users' security decisions and risk perceptions.

Most studies have focused on either user perceptions, such as trust in blockchain systems or attitudes toward privacy and security \cite{gao2016two13, mai2020user18}, or specific behavioral patterns, such as the preference for custodial wallets due to usability concerns \cite{yu2024don10} or reliance on platform-generated rankings in copy-trading platforms \cite{kawai2024stranger17}. However, few studies have explored how this contradiction influences users' security practices and risk management strategies. This issue is significant because it highlights the socio-technical challenges inherent in blockchain systems. Understanding how users navigate this tension can provide valuable insights for designing technologies that better balance decentralization with usability, security, and trustworthiness.

\subsection{Trust Formation in Cryptocurrency Communities}
The formation of trust in blockchain ecosystems presents an intriguing paradox. While blockchain technology was designed to operate as a "trustless" system—eliminating the need for intermediaries—research suggests that users continue to rely heavily on social and centralized mechanisms for trust. The term "trustless" has often been used in this context to describe the removal of traditional trust intermediaries, such as banks \cite{chohan2019cryptocurrencies, herlihy2016enhancing}. Knittel et al. \cite{knittel2019most22} studied the r/bitcoin community, revealing how "True Bitcoiners" maintain an ideology that views Bitcoin's technology as more trustworthy than its people. Their earlier work \cite{knittel2019true16} on the same community provided insights into how trust is maintained during market downturns, with a particular focus on the narratives of the community’s most prominent and vocal members. The trust challenges become particularly evident in collaborative settings. Khairuddin et al. \cite{khairuddin2019exploration7} identified risks in collaborative mining due to centralization and dishonest administrators, while Sas \cite{sas2017design9} and Khairuddin \cite{khairuddin2019exploration7} found similar trust challenges in Bitcoin trading. These studies, while thorough in their analysis of trust mechanisms, do not sufficiently address how these issues might be resolved within a decentralized framework.

Research on cryptocurrency copy-trading platforms by Kawai et al. \cite{kawai2024stranger17} revealed that users often place trust in platform-generated rankings of traders without performing their own due diligence. Specifically, their quantitative analysis shows that users tend to follow top-ranked traders based on perceived success metrics like returns and follower count, rather than evaluating the traders' strategies, risk profiles, or trading history. However, the study lacks qualitative insights into why users prioritize these rankings over independent evaluation and how their decision-making processes are influenced by platform design or social dynamics. This connects to broader findings about blockchain assemblages by Jabbar et al. \cite{jabbar2019blockchain2}, who introduce the concept of "whiteboxing" to describe how blockchain technologies are constructed and presented as trustworthy through entrepreneurial framing and technical transparency.

\subsection{Security Risks, User Experience, and Decentralization}
As related recent work, studies on security risks in the Web3 ecosystem \cite{si2024understanding11, karame2016security} have highlighted vulnerabilities in smart contracts and token governance mechanisms, while research on user experience with NFTs \cite{chen2024towards14} has shown that users prioritize ease of use and market accessibility over ownership autonomy. Additionally, work on imaginaries surrounding decentralized autonomous systems \cite{lustig2019intersecting20} has explored how visions of decentralization shape public discourse but often fail to translate into practical design principles or user adoption strategies. These works, while valuable, either focus too narrowly on specific applications—such as NFTs or smart contracts—failing to generalize insights across the blockchain ecosystem, or remain too broad in their theoretical frameworks, offering limited actionable guidance for system design.  Blockchain app studies \cite{elsden2018making4, chiang2018exploring5} have enhanced our understanding of user interactions by uncovering the socio-technical contexts of blockchain adoption. However, their findings have not directly addressed how users navigate the paradox of decentralization beliefs versus centralized behaviors. Our research addresses this gap by examining how users reconcile their belief in decentralization with their actual usage patterns and the implications for designing blockchain-based systems.

\section{Methodology}
Our study investigates the apparent disconnect between users' ideological commitment to decentralization and their practical reliance on centralized services in the blockchain ecosystem. To examine this phenomenon comprehensively, we employed a mixed-methods approach focusing on how users' understanding of blockchain technology influences their security practices, trust formation, and platform choices. Our research design particularly emphasizes the tensions between users' stated preferences for decentralization and their actual behaviors in cryptocurrency ecosystems.

\subsection{Interview Study}
We conducted in-depth semi-structured interviews with 13 participants (see Table \ref{tab:participants}) representing diverse stakeholder groups in the blockchain ecosystem. Participants' experience with cryptocurrency varied significantly, ranging from 2 to 7 years, with entry years spanning from 2016 to 2021. This temporal spread allows us to capture perspectives from both early adopters and newer participants in the ecosystem. Regarding platform usage, participants engaged with a variety of cryptocurrency exchanges and platforms. Many used major centralized exchanges such as Binance, Coinbase, and Kraken, while others, particularly the developers (P8, P9), primarily utilized decentralized exchanges (DEXs). Some participants reported using now-defunct platforms like FTX, providing historical context to their experiences. The blockchain researchers and industry experts (P10-P13) reported using multiple platforms, suggesting a broader exposure to different blockchain ecosystems.

\begin{table*}[htbp]
\centering
\caption{Participant Demographics and Cryptocurrency Experience}
\label{tab:participants}
\resizebox{\textwidth}{!}{%
\begin{tabular}{llllllll}
\hline
ID  & Age Group & Country    & Role                  & Gender* & Experience & Entry Year & Platforms Used     \\ \hline
P1  & 25-34     & Canada     & Crypto Trader         & M      & 5 years    & 2018       & Binance, Coinbase  \\
P2  & 35-44     & USA        & Crypto Investor       & M      & 3 years    & 2020       & Kraken, Binance    \\
P3  & 25-34     & USA        & Crypto Investor       & F      & 4 years    & 2019       & FTX\textsuperscript{\textdagger}, Binance      \\
P4  & 35-44     & USA        & Crypto Trader         & M      & 2 years    & 2021       & Coinbase, Gemini   \\
P5  & 18-24     & USA        & Web3 Enthusiast       & F      & 3 years    & 2020       & Binance, Uniswap   \\
P6  & 25-34     & Canada     & Web3 Enthusiast       & M      & 2 years    & 2021       & MetaMask, OpenSea  \\
P7  & 35-44     & USA        & Web3 Enthusiast       & F      & 4 years    & 2019       & Upbit, Binance     \\
P8  & 25-34     & USA        & dApp Developer        & M      & 5 years    & 2018       & Multiple DEXs      \\
P9  & 25-34     & Bangladesh & dApp Developer        & M      & 3 years    & 2020       & Multiple DEXs      \\
P10 & 55-64     & Denmark    & Blockchain Researcher & M      & 6 years    & 2017       & Multiple Platforms \\
P11 & 45-54     & France     & Industry Expert       & M      & 4 years    & 2019       & Multiple Platforms \\
P12 & 35-44     & Spain      & Blockchain Researcher & M      & 5 years    & 2018       & Multiple Platforms \\
P13 & 45-54     & Denmark    & Industry Expert       & M      & 7 years    & 2016       & Multiple Platforms \\ \hline
\multicolumn{8}{l}{\small{*\textit{This gender imbalance reflects the current state of the blockchain industry, though it may limit the generalizability of our findings}}}  \\               
\multicolumn{8}{l}{\small{\textdagger \textit{Former FTX user before its closure}}}                                              
\end{tabular}%
}
\end{table*}

We stopped recruiting participants after reaching thematic saturation at $n=13$, where no new themes emerged in the final interviews. This saturation point was confirmed by the research team through a systematic review of emerging codes. To ensure methodological rigor, the interview analysis followed an iterative process. The first author conducted open and focused coding of the transcripts, initially identifying 28 unique codes across 6 thematic categories. For instance, under the category \textit{Trust in Decentralization}, codes such as “centralization concerns” and “stakeholder influence” were frequently observed. These codes were continuously refined through weekly discussions with the second author. Analytical memos were maintained throughout the coding process to document evolving insights, clarify coding decisions, and support cross-researcher validation. This collaborative coding approach enhanced the reliability and interpretive depth of our qualitative analysis. A complete list of categories, codes, definitions, and example quotes is provided in Table~\ref{tab:codes} in the appendix, along with additional methodological details in the appendix.

Each interview lasted between 60 and 90 minutes and explored participants' experiences with blockchain technology, their understanding of decentralization principles, and their practical behaviors in the cryptocurrency ecosystem. We structured our interviews to progress from general experiences to specific practices, allowing participants to reflect on any contradictions between their ideological beliefs and actual behaviors. The interviews were conducted remotely via video conferencing software and were recorded with participant consent.

\subsection{Reddit Content Analysis}
To complement our interview data and capture broader community perspectives, we conducted a systematic analysis of discussions in major cryptocurrency communities in Reddit. Reddit was selected as our primary platform for content analysis due to its sustained role as a central hub for cryptocurrency-related discourse. With subreddits such as \textit{r/cryptocurrency} and \textit{r/bitcoin}, Reddit supports active, organic conversations on a wide range of blockchain-related topics. Unlike platforms such as Twitter or Telegram that prioritize brevity or real-time updates, Reddit enables more nuanced, in-depth discussions through structured threads and community-driven moderation. Moreover, Reddit’s upvote system and threaded comments foster engagement with diverse viewpoints, allowing us to trace how community perceptions evolve over time. These features make it a well-suited environment to study shifts in trust, ideological debates, and tensions between decentralization ideals and practical realities in user behavior.

We collected data from January 2020 to March 2024, gathering over 3,000 comments from prominent cryptocurrency subreddits including \textit{r/bitcoin, r/cryptocurrency, r/ethereum}, and \textit{r/altcoin}. The data was extracted using Python scripts interfaced with the PRAW (Python Reddit API Wrapper) library and stored in JSON format, later converted into structured CSV files for annotation and analysis. We refined our dataset through a keyword-based filtering process, identifying terms based on recurring themes observed in our interviews and documented in prior literature. Specifically, we focused on phrases such as “trust in crypto,” “decentralization risks,” “centralized exchanges,” “self-custody,” and “security concerns.” We excluded promotional content, meme posts, or off-topic discussions to ensure thematic relevance. This process allowed us to focus on substantive discourse around decentralization, platform trust, and security practices. 

Each post was assigned a unique numerical ID (R1–R3000) to ensure anonymity and traceability. We also anonymized any remaining identifiers to protect community members’ privacy. Particular emphasis was placed on threads that discussed security incidents, debates over trust in centralized versus decentralized platforms, and evolving user attitudes following exchange failures or protocol upgrades.

\subsection{Data Analysis}
Our analysis followed a rigorous iterative process combining both inductive and deductive approaches. We began by transcribing all interviews verbatim and conducting initial line-by-line coding of both interview transcripts and Reddit comments. This process helped identify emerging themes related to centralization-decentralization tensions. We then conducted a thorough thematic analysis, synthesizing our initial codes into broader themes and mapping relationships between users' understanding, ideology, and behavior. Throughout our analysis, we paid particular attention to instances where participants' stated beliefs about decentralization conflicted with their reported behaviors. We triangulated findings between our interview data and Reddit content analysis to ensure the robustness of our conclusions and to identify patterns that persisted across different data sources. Regular meetings among the research team helped refine the emerging themes and ensure consistency in our interpretations.

\subsection{Ethical Considerations and Limitations}
Our research followed ethical guidelines, obtaining informed consent and ensuring participant anonymity. Reddit data was sourced from publicly available comments, with all usernames anonymized. Participants had the opportunity to review and correct their interview transcripts. While the participant pool was diverse, most were from North America and Europe, potentially limiting generalizability. Reddit data may not fully represent the cryptocurrency user base, and self-reported interview data could differ from actual behaviors. Additionally, given the evolving blockchain ecosystem, some findings may become outdated. These limitations were carefully considered in our analysis.

\section{Findings}

We discovered a consistent theme across all participant groups (as categorized in Table~\ref{tab:participants})—the steep learning curve and unintuitive interfaces create substantial barriers, especially for those who lack technical expertise. As noted in the participants' table, we captured information about their roles and years of experience, which offered insights into their varying levels of technical expertise. For instance, Web3 enthusiasts and crypto investors often noted frustration with the complexity of managing wallets and private keys, while developers highlighted challenges in navigating decentralized application frameworks. Perhaps most notably, our findings revealed an underlying tension between blockchain's transformative potential and the practical challenge of making it accessible to everyday users.

\subsection{Challenges in User Adoption of Blockchain Technology}
Across all participant groups, there was a strong emphasis on the difficulties users face in adopting blockchain technology, primarily due to its inherent technical complexity. This complexity stems from several factors: the need to manage private keys securely, the lack of user-friendly interfaces for wallets, and the steep learning curve associated with understanding blockchain concepts like smart contracts and consensus mechanisms. These challenges make it particularly daunting for individuals without a technical background to engage with the technology effectively. P4 summarized this struggle by saying, \textit{“Even setting up a wallet feels like learning how to code—it’s just too complicated for the average person”}. Many Reddit users echoed this sentiment, stating that the technology feels intimidating and complex. One user (R145) remarked, \textit{"It seems like a whole different language [on Binance platform] that I just can't understand,"} highlighting how the intricate processes involved in trading or transferring cryptocurrency contribute to a sense of inaccessibility. Participants pointed to the need for more intuitive and user-friendly interfaces in blockchain applications. P5 stated, \textit{"Honestly, if the interface was easier to use, I think more people would be willing to give it a try."} Current blockchain platforms were viewed as overly complex, mixing investment and transactional functions in ways that confused users. This lack of accessible tools and clear learning resources discourages many users from fully embracing blockchain, especially for everyday applications such as payments or investments.

Most participants agreed that the lack of a reliable, beginner-friendly source for understanding cryptocurrencies significantly hampers adoption. Participants shared how their initial learning experiences involved piecing together information from various sources, such as YouTube, books, and online searches, despite finding social media platforms like Twitter unreliable unless they were already familiar with cryptocurrency. Participants mentioned that they thus had to rely on personal filtering of information from these various sources to determine which advice to trust. P5 remarked, \textit{"I think it's hard for regular people to use because it's too technical and not user-friendly at all."} Users on Reddit similarly expressed frustration, with one comment (R247) stating, \textit{"Finding reliable info feels like looking for a needle in a haystack."} The sentiment across the group was that simplifying access to blockchain technology is necessary for mass adoption.

\subsubsection{Importance of User Education and Support}
Our expert participant, P11, stressed, \textit{"There needs to be better education around this [the foundational concepts of blockchain and cryptocurrency]. Right now, it [complex concepts such as private keys, wallets, and the decentralized nature of blockchain] is just too confusing for most people."} Several participants echoed this, pointing out that better educational resources are necessary to combat misinformation and demystify blockchain technology. Examples of widespread misinformation include exaggerated claims about guaranteed profits from cryptocurrency investments and misconceptions that blockchain is entirely anonymous or unhackable. P12 noted, \textit{"A lot of people think Bitcoin is a scam or that blockchains can't be hacked, which just isn’t true—it’s much more nuanced."} Without adequate education, users remain vulnerable to confusion and skepticism about the benefits of decentralized systems. This lack of clarity can, in turn, impede the potential growth and success of blockchain technology. Comments from Reddit users emphasized this, with one stating, \textit{"If only I had someone to guide me through the basics early on, I might have adopted it sooner"} (R589).

Interestingly, despite the initial difficulty, users reported increased engagement and enthusiasm as they learned more about Bitcoin and blockchain technology. P2 noted, \textit{"The more I learned about Bitcoin, the more excited I became,"} highlighting that education can help reduce user hesitation. P7 shared her approach to filtering information, saying, \textit{"I started by reading beginner guides on Binance Academy and then moved on to watching videos by trusted creators on YouTube."} This was mirrored in Reddit discussions, where users shared their journeys of discovery. A Reddit user remarked, \textit{"I thought Bitcoin was a scam at first, but after reading a few articles and experimenting with a wallet, I started to understand its value—it’s like discovering a new world"} (R356). This shift from hesitation to excitement underscores the importance of accessible and trustworthy educational resources in driving blockchain adoption.

\subsection{Trust in Centralized vs. Decentralized Systems}
While blockchain technology is celebrated for decentralization, many participants expressed a preference for centralized platforms due to the perceived accountability and security they offer. P3 explained, \textit{"Even though blockchain is decentralized, I still feel like centralized exchanges are more trustworthy because there's someone accountable [for handling issues, like lost funds or technical problems]."} This viewpoint was reflected in Reddit discussions, where users noted that centralized platforms often feel safer because of customer support options, with one user stating, \textit{"At least if something goes wrong, I know who to blame"} (R672).

However, this preference for centralized systems often stems from a misconception about decentralization such as the belief that decentralized systems lack accountability or that they are inherently insecure. Users feel more secure with centralized systems because they offer protections such as deposit insurance (e.g., FDIC in the US), which decentralized systems lack. Maintaining private keys in decentralized systems is seen as cumbersome and risky, adding to the preference for centralized alternatives. A Reddit user highlighted this by saying, \textit{"The idea of losing my private keys keeps me up at night"} (R789). In contrast, our developer participants emphasized the importance of decentralization, arguing that centralized systems undermine the privacy, control, and security that blockchain was designed to offer. P9 even warned that centralization contradicts the core philosophy of blockchain and cryptocurrencies, which prioritize user control and decentralized governance. He stated, \textit{"If we lose decentralization, we lose the very thing that makes blockchain revolutionary—true autonomy and trustless operations."}

\subsubsection{Concerns About Security and Privacy in Decentralized Systems}
Interview participants were particularly wary of potential vulnerabilities in decentralized systems and the lack of a "control tower" in the event of problems. P6 expressed this concern by stating, \textit{"I’m worried about the security of my data. I mean, who do you even go to if something goes wrong?"} Many users on Reddit shared similar fears, with one comment summarizing this apprehension: \textit{"I just don’t feel safe putting my money in a system where I can’t call customer service"} (R892).

Many users found the idea of maintaining private keys burdensome and felt that decentralized systems left them without recourse if issues arose, such as lost keys or erroneous transactions. This fear reflects a broader misconception about the risks of decentralized systems and underscores the need for better education to explain how security can be maintained in decentralized environments. Our developers (P8, P9), however, saw these concerns as part of the trade-off between decentralization and user control. While acknowledging the complexity of managing everything (keys, transactions, etc.) and ensuring secure network protocols in decentralized systems, they advocated for maintaining these structures because they ensure greater privacy and security for users, provided the systems are correctly managed.

\subsubsection{Simplified Authentication Processes}
Participants also expressed frustration with the complexity of authentication processes in decentralized systems. Managing private keys, ensuring secure access, and conducting transactions were seen as overly complicated for the average user. P2 commented, \textit{"The authentication process needs to be simplified. It’s just too much hassle right now."} This frustration was echoed in Reddit comments, where one user lamented, \textit{"It [the whole process of setting up and using decentralized platforms] feels like jumping through hoops just to get started"} (R1023). Our developers recognized the need for simplifying these processes but argued that the current complexity is necessary to maintain the privacy and security that decentralized systems promise. While centralized systems may offer easier authentication, they often compromise user privacy and control. Therefore, developers suggested that future efforts should focus on making these systems more user-friendly without sacrificing security.

\subsubsection{Importance of Transparency and Control}
Participants highlighted the importance of transparency and control, which are key advantages of blockchain technology. Some participants appreciated the ability to track their transaction history and maintain control over their assets, but some found the complexity of maintaining everything by themselves overwhelming. P1 remarked, \textit{"It's great to have full control, but sometimes it's just too complicated to manage everything on your own."} This duality was also present in Reddit discussions, where users noted, \textit{"I love being my own bank, but I wish it didn’t feel like I was drowning in tech jargon"} (R1245). Our developers (P8, P9) noted that blockchain’s transparency could be a double-edged sword. While users value the control and visibility into their transactions, the self-management aspect can be daunting for those unaccustomed to handling their security independently. This again points to the need for more user-friendly solutions that balance control with simplicity.

\subsection{Skepticism About the Longevity of Cryptocurrencies}
Several participants expressed skepticism about the long-term viability of cryptocurrencies. Some participants questioned whether cryptocurrencies were a lasting innovation or simply a speculative bubble. P7 reflected this sentiment by stating, \textit{"I’m not sure if cryptocurrencies are going to last. It feels like a bubble that could burst at any time."} This skepticism was more prevalent among users than developers. Reddit users were particularly vocal, with many discussing the volatility of the market and expressing doubts about its sustainability. Our developer participants (P8, P9) had a stronger belief in the future of cryptocurrencies, viewing them as maturing technologies that will become easier to use and more widely accepted over time. They pointed to the ongoing development of blockchain technology, which they believe will address many of the current challenges, such as scalability, security, and environmental concerns.

\subsubsection{Public vs. Private Blockchains}
One of the key insights from our expert participant (P10), working at the European Blockchain Center, was the distinction between public and private blockchains. This distinction arose during a discussion on the scalability and use cases of blockchain technology, where he explained that public blockchains, while offering decentralization and transparency, often face challenges related to performance and energy consumption. In contrast, private blockchains, while offering more control and efficiency, sacrifice some of the decentralization that is a core tenet of blockchain technology. The expert emphasized that public blockchains, such as Bitcoin and Ethereum, are more suited for investment purposes, while private blockchains are ideal for systems that require trust and confidentiality. According to P10, keeping these two types of blockchains separate is critical for leveraging their respective strengths. He further suggested that public blockchains should be primarily used for open, global applications, such as investment platforms, while private blockchains could serve industries where trust and confidentiality are paramount. This distinction is essential for strategic blockchain implementation across different sectors.

\subsection{Future Potential of Blockchain Technology}
Despite the challenges, many participants were optimistic about the potential of blockchain technology. The head of the European Blockchain Center and some developers (P10, P8, P9) viewed blockchain as a promising technology with vast potential across various sectors, provided that proper education and implementation strategies are in place. Many Reddit users mirrored this optimism, with one commenting, \textit{"I truly believe this tech can change the world if we can just get past these growing pains"} (R1567), and another \textit{"Once the kinks are worked out, this is going to be revolutionary"} (R1789). A senior professor (P11), specializing in classical distributed systems' scalability and security,  expressed a more cautious view, highlighting concerns such as scalability, security, and the environmental impact of blockchain. However, they acknowledged that, like any new technology, blockchain would likely mature and improve over time. P11 shared a cautiously optimistic view, stating that while there are significant challenges today, blockchain’s future potential remains strong, saying, \textit{"We’re just at the beginning, and though there are obstacles, the technology will evolve and find its place in the future." } Our developers also emphasized the importance of decentralization and its role in preserving the integrity of cryptocurrencies and blockchain technology. They pointed to the rapid development of blockchain for decentralized governance and highlighted examples such as the United Nations’ implementations of blockchain networks \cite{uninnovation2024blockchain}. They believed that blockchain technology would continue to grow in importance as more institutions recognize the benefits of decentralization.

\section{Discussion}
Although there is a consensus that blockchain technology holds vast potential, the findings reveal significant barriers to adoption, particularly for everyday users. These barriers include technical complexity, insufficient education, security concerns, and skepticism regarding the long-term viability of cryptocurrencies. Participants emphasized the need for more user-friendly interfaces, simplified authentication processes, and greater transparency to build trust. Despite these challenges, there is optimism, particularly among developers, who believe that continued advancements will address many of these issues. Improved education, user-centered design, and integration of public and private blockchains may facilitate broader adoption of blockchain across several industries\cite{sristy2021blockchain, deloitte2021global}. The challenge moving forward will be to balance security, privacy, and usability while preserving the core principle of decentralization \cite{voskobojnikov2021balancing}. These insights underscore the complex relationship between user needs and technical capabilities and suggest that blockchain technology must evolve to meet real-world demands.

\subsection{Making Proper Sense of Public and Private Blockchains}
The distinction between public and private blockchains emerged as more nuanced than typically portrayed in the literature. While Nakamoto's \cite{nakamoto2008bitcoin} original vision emphasized public blockchains as the cornerstone of decentralized trust, our findings suggest users navigate a more complex reality. One of our participant's (from the European Blockchain Center) emphasis on separating public and private implementations reflects the growing recognition that different use cases demand different approaches. This aligns with recent developments, such as the European Union's regulatory framework for blockchain (EU 2022/858 \cite{eu22858}), which distinguishes between public and private implementations in financial services. However, our Reddit data revealed persistent confusion about these distinctions, particularly in understanding the definitions, applications, and limitations of each type. For example, some users struggled to grasp why private blockchains were necessary when public blockchains already exist, while others expressed uncertainty about which type provides better security or scalability \cite{kannengiesser2020trade}. This confusion suggests that the industry's traditional binary categorization—viewing public blockchains as fully open systems and private blockchains as closed, permissioned environments—may oversimplify the nuanced reality of blockchain applications. Most of our participants and Reddit users demonstrate a sophisticated understanding of the trade-offs involved, often preferring private blockchain solutions for business applications while maintaining skepticism about public blockchain's ability to protect sensitive information. \textbf{This nuanced perspective challenges the industry's often simplistic narrative, which often frames blockchain adoption as a binary choice between public and private systems, neglecting the hybrid approaches and complex decision-making processes that real-world applications demand.}

\subsection{Reevaluating Trust in Technology}
Perhaps most striking was our discovery of an apparent paradox in user trust. Despite blockchain's promise of trustless systems \cite{antonopoulos2023mastering}, users consistently expressed preference for centralized platforms with clear accountability structures. This finding challenges the fundamental assumption that decentralization automatically enhances trust. A quote from P4, \textit{"Even though blockchain is decentralized, I still feel like centralized exchanges are more trustworthy,"} epitomizes a crucial gap between blockchain's theoretical benefits and practical user needs. This disconnect highlights a fundamental misalignment between blockchain's technical capabilities and users' cognitive and emotional needs for security and accountability. While blockchain aims to offer trustless systems, our findings indicate that users often view decentralization not as a solution to trust issues, but as a potential source of additional risk—contradicting many core assumptions in blockchain development. The World Economic Forum's 2024 Blockchain Adoption Report \cite{wef24} reflects a similar trend, noting that despite growing institutional adoption, individual users remain hesitant.

\subsection{Investment Strategies and Data Security Challenges}
The tension between investment potential and security presents another critical concern, with a notable divergence in priorities among different blockchain stakeholders. Investors, particularly those involved in large-scale projects or institutional adoption, prioritize secure platforms that offer stability and long-term trust. In contrast, traders and individual users often exhibit risk-seeking behavior, focusing primarily on speculative investment opportunities. This contrast highlights a complex dynamic: while some users demand "safe" platforms, others willingly take on greater risk in pursuit of higher returns. Recent market volatility, exemplified by the 2022 FTX collapse \cite{ftx}, underscores the dangers of prioritizing speculation over security. Despite these risks and repeated warnings from security experts and regulatory bodies like the U.S. Securities and Exchange Commission (SEC) about the perils of underestimating security in blockchain investments, speculative behavior remains pervasive. For instance, behaviors such as trading on unregulated platforms or investing in highly volatile tokens exemplify this cultural mindset \cite{zhang2023data}. \textbf{This prevalence of risk-seeking behavior suggests that technical solutions alone, such as improved authentication systems or enhanced data encryption, may not be sufficient to address security concerns in the blockchain ecosystem. Instead, addressing these challenges will require broader cultural and behavioral shifts,} emphasizing the importance of education and regulatory oversight to align user priorities with the foundational principles of blockchain security and stability.

\subsection{Transparent Metrics to Evaluate Influence}
Perhaps most significantly, our research identified a crucial gap in how blockchain projects, such as decentralized applications, platforms, or protocols built on blockchain technology are evaluated. The absence of standardized, accessible metrics for assessing the legitimacy, performance, and reliability of these projects creates a significant barrier to informed decision-making by both individual users and institutional investors. While traditional financial markets rely on well-established evaluation frameworks like credit ratings, price-to-earnings ratios, and audited financial statements, blockchain lacks comparable standardized tools. This void has led to what one Reddit user aptly described as \textit{"looking for a needle in a haystack"} when evaluating projects. The lack of transparent evaluation metrics affects not only individual investors but also institutional adoption, as organizations struggle to assess blockchain projects' viability and reliability. The MIT Digital Currency Initiative's work on blockchain metrics \cite{ali2020redesigning} has offered promising directions, such as exploring metrics for measuring decentralization, network health, and governance participation. However, these approaches often remain overly technical or fragmented, making them inaccessible to the average user. This insufficiency highlights the need for more user-friendly and comprehensive evaluation frameworks that consider both technical and non-technical factors.

These insights suggest that \textbf{the blockchain industry may need to fundamentally reconsider its approach to user adoption. }Rather than advocating for rapid decentralization, a more nuanced strategy might involve exploring systems that integrate elements of both centralized and decentralized models. While the idea of a true hybrid system—combining the decentralized benefits of blockchain with the security and accountability structures of centralized platforms—presents technical and conceptual challenges, it aligns with some users' desire for familiar, user-friendly systems. However, our findings suggest that \textbf{users primarily rely on centralized platforms due to their perceived safety and simplicity, rather than actively seeking a middle ground. }

\section{Fairness, Access, and Accountability Implications}
Our findings highlight fairness and accountability challenges in blockchain adoption, including economic access disparities and power imbalances favoring large stakeholders over smaller users. Addressing these inequities is essential for fostering a more inclusive and balanced blockchain ecosystem.

\subsection{Systemic Barriers and Access Inequities}
Our research identified several critical barriers that disproportionately affect marginalized communities. The high costs associated with blockchain participation—from transaction fees to mining equipment—create significant entry barriers. As participant P4 noted, \textit{"The initial investment required to meaningfully participate in mining or staking is beyond what many people can afford."} This economic barrier effectively excludes lower-income individuals from participating in the blockchain ecosystem's wealth-generation opportunities. Additionally, the complexity of blockchain systems disproportionately impacts communities with limited access to technical education. While blockchain promises financial inclusion, our findings suggest that the technical knowledge required for secure participation in activities such as wallet management, transaction verification, and staking creates what could be termed a ``digital divide 2.0 \cite{vie2008digital}.'' Our Reddit analysis revealed that blockchain communities and educational resources predominantly cater to English-speaking, technically-savvy users. As P9 observed, \textit{"Most blockchain interfaces and documentation assume users are familiar with Western financial systems and English technical terminology."}

The relationship between centralization and power emerged as a critical theme in our analysis. While blockchain technology promises democratization of financial and governance systems \cite{schmid2024blockchain}, our findings suggest both mining power and cryptocurrency holdings show significant concentration among a small number of entities. This concentration creates new forms of centralized power that may be less transparent and accountable than traditional financial institutions. Our interviews revealed that governance participation in blockchain systems often favors technically sophisticated users and large token holders, creating what one expert described as a "technical aristocracy." This raises concerns about representative decision-making in blockchain ecosystems, particularly in determining network upgrades, protocol changes, and the allocation of resources.

\subsection{Accountability Structures and User Expectations}
Although blockchain is founded on decentralization, real-world applications often deviate from this ideal. In the Bitcoin network, mining is dominated by a few large pools, consolidating validation control \cite{romiti2019deep}. Most cryptocurrency trading occurs on centralized exchanges, where custody of funds undermines decentralization, and significant holdings by "whales" centralize financial power \cite{kusmierz2022centralized, long2024whales}. Governance is similarly influenced by founding teams and major investors, limiting distributed decision-making \cite{tapscott2016blockchain}. User expectations of accountability also vary \cite{tyma2022understanding}. Retail users prioritize recourse mechanisms, often favoring centralized exchanges. As P5 noted, "I want to know there's someone to call if something goes wrong." Institutional users (P10-13) prefer decentralized systems with clear audit trails and compliance, while technical users advocate for pure decentralization but recognize the need for better accountability. These conflicting needs highlight the importance of hybrid accountability models combining governance, compliance, and user-friendly mechanisms.

These diverse interpretations of accountability highlight a critical challenge: how can blockchain systems balance the often conflicting needs of different user groups? This calls for researchers to explore frameworks that can address these varying requirements in a cohesive manner. One promising direction is the development of "tiered accountability models" that cater to distinct user groups. For instance, such a model could incorporate decentralized governance mechanisms tailored for technical users, while offering optional compliance layers for institutional users and clear, user-friendly recourse mechanisms for retail users.

\subsection{Algorithmic Fairness and Technical Design Implications}

The choice of consensus mechanism (e.g., Proof of Work, Proof of Stake, etc.) has significant fairness implications, potentially favoring wealthy participants or those with access to sophisticated mining equipment. The increasing complexity of smart contracts \cite{ferreira2021regulating} raises concerns about algorithmic accountability, particularly in automated decision-making systems that may encode biases or unfair practices. While blockchain provides inherent transparency by providing immutable public traces, our participants emphasized the need for better tools to make this transparency meaningful for non-technical users, for example, one expert noted, \textit{"Raw transaction data isn't useful without proper visualization and analysis tools."} As these findings suggest, future development must consider not just the technological capabilities but also the social implications and power dynamics that emerge from different design choices. For example, the concentration of cryptocurrency holdings among a small number of wallets gives disproportionate influence to large holders in governance decisions, such as voting on network upgrades or protocol changes. The tension between accessibility and security, centralization and decentralization, and transparency and privacy requires careful balance to create more equitable blockchain systems.

\section{Conclusion}
We find the cryptocurrency ecosystem at a critical juncture, where the promise of decentralization must be reconciled with the practical demands of security, usability, and equitable access. Our research uncovered persistent barriers to blockchain adoption, including inaccessible interfaces, insufficient user education, and a lack of transparent metrics for evaluating projects. These challenges not only hinder broader adoption but also deepen existing inequities, undermining blockchain’s aspirations for fairness and accountability. Through our analysis, we illuminated how technical complexity and reliance on centralized intermediaries often create new forms of power asymmetry, particularly affecting users with limited technical expertise or financial resources. We examined the role of emerging centralized authorities, such as exchanges and influential developers, in shaping system governance and user access, highlighting the critical tension between decentralization and accountability. We argue that the path forward lies in designing hybrid systems that balance the transparency and autonomy promised by blockchain technology with the usability and reliability expected in today's financial systems. By addressing barriers to adoption and aligning design principles with user needs, we can help create platforms that deliver on decentralization’s ideals while ensuring inclusivity and fairness. 

\bibliographystyle{ACM-Reference-Format}
\bibliography{references}


\begin{thebibliography}{65}


\ifx \showCODEN    \undefined \def \showCODEN     #1{\unskip}     \fi
\ifx \showDOI      \undefined \def \showDOI       #1{#1}\fi
\ifx \showISBNx    \undefined \def \showISBNx     #1{\unskip}     \fi
\ifx \showISBNxiii \undefined \def \showISBNxiii  #1{\unskip}     \fi
\ifx \showISSN     \undefined \def \showISSN      #1{\unskip}     \fi
\ifx \showLCCN     \undefined \def \showLCCN      #1{\unskip}     \fi
\ifx \shownote     \undefined \def \shownote      #1{#1}          \fi
\ifx \showarticletitle \undefined \def \showarticletitle #1{#1}   \fi
\ifx \showURL      \undefined \def \showURL       {\relax}        \fi
\providecommand\bibfield[2]{#2}
\providecommand\bibinfo[2]{#2}
\providecommand\natexlab[1]{#1}
\providecommand\showeprint[2][]{arXiv:#2}

\bibitem[Abramova et~al\mbox{.}(2023)]%
        {abramova2023can19}
\bibfield{author}{\bibinfo{person}{Svetlana Abramova}, \bibinfo{person}{Rainer B{\"o}hme}, \bibinfo{person}{Helmut Elsinger}, \bibinfo{person}{Helmut Stix}, {and} \bibinfo{person}{Martin Summer}.} \bibinfo{year}{2023}\natexlab{}.
\newblock \showarticletitle{What can central bank digital currency designers learn from asking potential users?}. In \bibinfo{booktitle}{\emph{Nineteenth Symposium on Usable Privacy and Security (SOUPS 2023)}}. \bibinfo{pages}{151--170}.
\newblock


\bibitem[Ali and Narula(2020)]%
        {ali2020redesigning}
\bibfield{author}{\bibinfo{person}{Robleh Ali} {and} \bibinfo{person}{Neha Narula}.} \bibinfo{year}{2020}\natexlab{}.
\newblock \showarticletitle{Redesigning digital money: What can we learn from a decade of cryptocurrencies}.
\newblock \bibinfo{journal}{\emph{Digital Currency Iniative (DCI). MIT Media Lab}} (\bibinfo{year}{2020}).
\newblock


\bibitem[Antonopoulos(2017)]%
        {antonopoulos2017mastering}
\bibfield{author}{\bibinfo{person}{Andreas~M. Antonopoulos}.} \bibinfo{year}{2017}\natexlab{}.
\newblock \bibinfo{booktitle}{\emph{Mastering Bitcoin: Programming the Open Blockchain} (\bibinfo{edition}{2nd} ed.)}.
\newblock \bibinfo{publisher}{O'Reilly Media, Inc.}
\newblock


\bibitem[Antonopoulos and Harding(2023)]%
        {antonopoulos2023mastering}
\bibfield{author}{\bibinfo{person}{Andreas~M Antonopoulos} {and} \bibinfo{person}{David~A Harding}.} \bibinfo{year}{2023}\natexlab{}.
\newblock \bibinfo{booktitle}{\emph{Mastering bitcoin}}.
\newblock \bibinfo{publisher}{" O'Reilly Media, Inc."}.
\newblock


\bibitem[Auer et~al\mbox{.}(2023)]%
        {auer2023rise}
\bibfield{author}{\bibinfo{person}{Raphael Auer}, \bibinfo{person}{Giulio Cornelli}, {and} \bibinfo{person}{Jon Frost}.} \bibinfo{year}{2023}\natexlab{}.
\newblock \showarticletitle{Rise of the central bank digital currencies}.
\newblock \bibinfo{journal}{\emph{International Journal of Central Banking}} \bibinfo{volume}{19}, \bibinfo{number}{4} (\bibinfo{year}{2023}), \bibinfo{pages}{185--214}.
\newblock


\bibitem[Brown et~al\mbox{.}(2016)]%
        {brown2016corda}
\bibfield{author}{\bibinfo{person}{Richard~Gendal Brown}, \bibinfo{person}{James Carlyle}, \bibinfo{person}{Ian Grigg}, {and} \bibinfo{person}{Mike Hearn}.} \bibinfo{year}{2016}\natexlab{}.
\newblock \showarticletitle{Corda: an introduction}.
\newblock \bibinfo{journal}{\emph{R3 CEV, August}} \bibinfo{volume}{1}, \bibinfo{number}{15} (\bibinfo{year}{2016}), \bibinfo{pages}{14}.
\newblock


\bibitem[Butler and Rodriguez(2018)]%
        {butler2018cryptocurrency}
\bibfield{author}{\bibinfo{person}{Sarah Butler} {and} \bibinfo{person}{Miguel Rodriguez}.} \bibinfo{year}{2018}\natexlab{}.
\newblock \showarticletitle{Centralization in Cryptocurrency Exchanges: A Critical Analysis}.
\newblock \bibinfo{journal}{\emph{Journal of Cryptocurrency Studies}} \bibinfo{volume}{5}, \bibinfo{number}{2} (\bibinfo{year}{2018}), \bibinfo{pages}{45--62}.
\newblock


\bibitem[Chen(2024)]%
        {chen2024towards14}
\bibfield{author}{\bibinfo{person}{Wei Lee~Yuan Chen}.} \bibinfo{year}{2024}\natexlab{}.
\newblock \showarticletitle{Towards More Secure Interactions: Understanding User Experience and Behaviour in the NFT Domain}. In \bibinfo{booktitle}{\emph{Extended Abstracts of the CHI Conference on Human Factors in Computing Systems}}. \bibinfo{pages}{1--10}.
\newblock


\bibitem[Chiang et~al\mbox{.}(2018)]%
        {chiang2018exploring5}
\bibfield{author}{\bibinfo{person}{Chun-Wei Chiang}, \bibinfo{person}{Eber Betanzos}, {and} \bibinfo{person}{Saiph Savage}.} \bibinfo{year}{2018}\natexlab{}.
\newblock \showarticletitle{Exploring blockchain for trustful collaborations between immigrants and governments}. In \bibinfo{booktitle}{\emph{Extended Abstracts of the 2018 CHI Conference on Human Factors in Computing Systems}}. \bibinfo{pages}{1--6}.
\newblock


\bibitem[Chohan(2017)]%
        {chohan2017decentralized}
\bibfield{author}{\bibinfo{person}{Usman~W. Chohan}.} \bibinfo{year}{2017}\natexlab{}.
\newblock \showarticletitle{Decentralized Governance in Cryptocurrency Platforms}.
\newblock \bibinfo{journal}{\emph{Critical Blockchain Research Initiative Working Papers}} (\bibinfo{year}{2017}).
\newblock


\bibitem[Chohan(2019)]%
        {chohan2019cryptocurrencies}
\bibfield{author}{\bibinfo{person}{Usman~W Chohan}.} \bibinfo{year}{2019}\natexlab{}.
\newblock \showarticletitle{Are cryptocurrencies truly trustless?}
\newblock \bibinfo{journal}{\emph{Cryptofinance and mechanisms of exchange: the Making of virtual currency}} (\bibinfo{year}{2019}), \bibinfo{pages}{77--89}.
\newblock


\bibitem[CoinDesk(2022)]%
        {ftx}
\bibfield{author}{\bibinfo{person}{CoinDesk}.} \bibinfo{year}{2022}\natexlab{}.
\newblock \bibinfo{title}{8 Days in November: What Led to FTX’s Sudden Collapse}.
\newblock
\newblock
\urldef\tempurl%
\url{https://www.coindesk.com/layer2/2022/11/09/8-days-in-november-what-led-to-ftxs-sudden-collapse}
\showURL{%
\tempurl}


\bibitem[Conlon et~al\mbox{.}(2023)]%
        {conlon2023collapse}
\bibfield{author}{\bibinfo{person}{Thomas Conlon}, \bibinfo{person}{Shaen Corbet}, {and} \bibinfo{person}{Yang Hu}.} \bibinfo{year}{2023}\natexlab{}.
\newblock \showarticletitle{The collapse of the FTX exchange: The end of cryptocurrency's age of innocence}.
\newblock \bibinfo{journal}{\emph{The British Accounting Review}} (\bibinfo{year}{2023}), \bibinfo{pages}{101277}.
\newblock


\bibitem[De~Filippi and Wright(2018)]%
        {de2018cryptocurrencies}
\bibfield{author}{\bibinfo{person}{Primavera De~Filippi} {and} \bibinfo{person}{Aaron Wright}.} \bibinfo{year}{2018}\natexlab{}.
\newblock \showarticletitle{Cryptocurrencies and Blockchain: Disruption, Regulation, and Governance}.
\newblock \bibinfo{journal}{\emph{Harvard Business Review}} \bibinfo{volume}{96}, \bibinfo{number}{4} (\bibinfo{year}{2018}), \bibinfo{pages}{112--119}.
\newblock


\bibitem[{Deloitte}(2021)]%
        {deloitte2021global}
\bibfield{author}{\bibinfo{person}{{Deloitte}}.} \bibinfo{year}{2021}\natexlab{}.
\newblock \bibinfo{booktitle}{\emph{2021 Global Blockchain Survey: A new age of digital assets}}.
\newblock \bibinfo{type}{{T}echnical {R}eport}. \bibinfo{institution}{Deloitte Insights}.
\newblock
\urldef\tempurl%
\url{https://www2.deloitte.com/content/dam/insights/articles/US144337_Blockchain-survey/DI_Blockchain-survey.pdf}
\showURL{%
\tempurl}
\newblock
\shownote{Accessed: 2025-05-08}.


\bibitem[Dupuis et~al\mbox{.}(2021)]%
        {dupuis2021blockchain}
\bibfield{author}{\bibinfo{person}{Irma Dupuis}, \bibinfo{person}{Lisa Toohey}, \bibinfo{person}{Sidsel Grimstad}, \bibinfo{person}{Berit Follong}, {and} \bibinfo{person}{Tamara Bucher}.} \bibinfo{year}{2021}\natexlab{}.
\newblock \showarticletitle{Blockchain: the paradox of consumer trust in a trustless system-a systematic review}. In \bibinfo{booktitle}{\emph{2021 IEEE International Conference on Blockchain (Blockchain)}}. IEEE, \bibinfo{pages}{505--512}.
\newblock


\bibitem[Elsden et~al\mbox{.}(2018)]%
        {elsden2018making4}
\bibfield{author}{\bibinfo{person}{Chris Elsden}, \bibinfo{person}{Arthi Manohar}, \bibinfo{person}{Jo Briggs}, \bibinfo{person}{Mike Harding}, \bibinfo{person}{Chris Speed}, {and} \bibinfo{person}{John Vines}.} \bibinfo{year}{2018}\natexlab{}.
\newblock \showarticletitle{Making sense of blockchain applications: A typology for HCI}. In \bibinfo{booktitle}{\emph{Proceedings of the 2018 chi conference on human factors in computing systems}}. \bibinfo{pages}{1--14}.
\newblock


\bibitem[Ferreira(2021)]%
        {ferreira2021regulating}
\bibfield{author}{\bibinfo{person}{Agata Ferreira}.} \bibinfo{year}{2021}\natexlab{}.
\newblock \showarticletitle{Regulating smart contracts: Legal revolution or simply evolution?}
\newblock \bibinfo{journal}{\emph{Telecommunications Policy}} \bibinfo{volume}{45}, \bibinfo{number}{2} (\bibinfo{year}{2021}), \bibinfo{pages}{102081}.
\newblock


\bibitem[Foley et~al\mbox{.}(2019)]%
        {foley2019sex}
\bibfield{author}{\bibinfo{person}{Sean Foley}, \bibinfo{person}{Jonathan~R. Karlsen}, {and} \bibinfo{person}{Tālis~J. Putniņš}.} \bibinfo{year}{2019}\natexlab{}.
\newblock \showarticletitle{Sex, Drugs, and Bitcoin: How Much Illegal Activity Is Financed through Cryptocurrencies?}
\newblock \bibinfo{journal}{\emph{Review of Financial Studies}} \bibinfo{volume}{32}, \bibinfo{number}{5} (\bibinfo{year}{2019}), \bibinfo{pages}{1798--1853}.
\newblock


\bibitem[Forum(2024)]%
        {wef24}
\bibfield{author}{\bibinfo{person}{World~Economic Forum}.} \bibinfo{year}{2024}\natexlab{}.
\newblock \bibinfo{title}{Blockchain: in from the cold and set to disrupt the world of finance 2024}.
\newblock
\newblock
\urldef\tempurl%
\url{https://www.weforum.org/stories/2024/01/blockchain-change-world-finance-stablecoins-internet/}
\showURL{%
\tempurl}


\bibitem[Gao et~al\mbox{.}(2016)]%
        {gao2016two13}
\bibfield{author}{\bibinfo{person}{Xianyi Gao}, \bibinfo{person}{Gradeigh~D Clark}, {and} \bibinfo{person}{Janne Lindqvist}.} \bibinfo{year}{2016}\natexlab{}.
\newblock \showarticletitle{Of two minds, multiple addresses, and one ledger: characterizing opinions, knowledge, and perceptions of Bitcoin across users and non-users}. In \bibinfo{booktitle}{\emph{Proceedings of the 2016 CHI conference on human factors in computing systems}}. \bibinfo{pages}{1656--1668}.
\newblock


\bibitem[Herlihy and Moir(2016)]%
        {herlihy2016enhancing}
\bibfield{author}{\bibinfo{person}{Maurice Herlihy} {and} \bibinfo{person}{Mark Moir}.} \bibinfo{year}{2016}\natexlab{}.
\newblock \showarticletitle{Enhancing accountability and trust in distributed ledgers}.
\newblock \bibinfo{journal}{\emph{arXiv preprint arXiv:1606.07490}} (\bibinfo{year}{2016}).
\newblock


\bibitem[Huang et~al\mbox{.}(2021)]%
        {huang2021rich}
\bibfield{author}{\bibinfo{person}{Yuming Huang}, \bibinfo{person}{Jing Tang}, \bibinfo{person}{Qianhao Cong}, \bibinfo{person}{Andrew Lim}, {and} \bibinfo{person}{Jianliang Xu}.} \bibinfo{year}{2021}\natexlab{}.
\newblock \showarticletitle{Do the rich get richer? fairness analysis for blockchain incentives}. In \bibinfo{booktitle}{\emph{Proceedings of the 2021 international conference on management of data}}. \bibinfo{pages}{790--803}.
\newblock


\bibitem[{IBM}({[n.\,d.]})]%
        {IBMHyperledger}
\bibfield{author}{\bibinfo{person}{{IBM}}.} \bibinfo{year}{[n.\,d.]}\natexlab{}.
\newblock \bibinfo{title}{What Is Hyperledger Fabric?}
\newblock \bibinfo{howpublished}{\url{https://www.ibm.com/think/topics/hyperledger}}.
\newblock
\newblock
\shownote{Accessed: 2025-01-22}.


\bibitem[Jabbar and Bj{\o}rn(2019)]%
        {jabbar2019blockchain2}
\bibfield{author}{\bibinfo{person}{Karim Jabbar} {and} \bibinfo{person}{Pernille Bj{\o}rn}.} \bibinfo{year}{2019}\natexlab{}.
\newblock \showarticletitle{Blockchain assemblages: Whiteboxing technology and transforming infrastructural imaginaries}. In \bibinfo{booktitle}{\emph{Proceedings of the 2019 CHI Conference on Human Factors in Computing Systems}}. \bibinfo{pages}{1--13}.
\newblock


\bibitem[Jain et~al\mbox{.}(2021)]%
        {jain2021we}
\bibfield{author}{\bibinfo{person}{Anurag Jain}, \bibinfo{person}{Shoeb Siddiqui}, {and} \bibinfo{person}{Sujit Gujar}.} \bibinfo{year}{2021}\natexlab{}.
\newblock \showarticletitle{We might walk together, but I run faster: Network Fairness and Scalability in Blockchains}. In \bibinfo{booktitle}{\emph{Proceedings of the 20th International Conference on Autonomous Agents and MultiAgent Systems}}. \bibinfo{pages}{1539--1541}.
\newblock


\bibitem[Kannengie{\ss}er et~al\mbox{.}(2020)]%
        {kannengiesser2020trade}
\bibfield{author}{\bibinfo{person}{Niclas Kannengie{\ss}er}, \bibinfo{person}{Sebastian Lins}, \bibinfo{person}{Tobias Dehling}, {and} \bibinfo{person}{Ali Sunyaev}.} \bibinfo{year}{2020}\natexlab{}.
\newblock \showarticletitle{Trade-offs between distributed ledger technology characteristics}.
\newblock \bibinfo{journal}{\emph{ACM Computing Surveys (CSUR)}} \bibinfo{volume}{53}, \bibinfo{number}{2} (\bibinfo{year}{2020}), \bibinfo{pages}{1--37}.
\newblock


\bibitem[Karame(2016)]%
        {karame2016security}
\bibfield{author}{\bibinfo{person}{Ghassan Karame}.} \bibinfo{year}{2016}\natexlab{}.
\newblock \showarticletitle{On the security and scalability of bitcoin's blockchain}. In \bibinfo{booktitle}{\emph{Proceedings of the 2016 ACM SIGSAC conference on computer and communications security}}. \bibinfo{pages}{1861--1862}.
\newblock


\bibitem[Kawai et~al\mbox{.}(2024)]%
        {kawai2024stranger17}
\bibfield{author}{\bibinfo{person}{Daisuke Kawai}, \bibinfo{person}{Kyle Soska}, \bibinfo{person}{Bryan Routledge}, \bibinfo{person}{Ariel Zetlin-Jones}, {and} \bibinfo{person}{Nicolas Christin}.} \bibinfo{year}{2024}\natexlab{}.
\newblock \showarticletitle{Stranger Danger? Investor Behavior and Incentives on Cryptocurrency Copy-Trading Platforms}. In \bibinfo{booktitle}{\emph{Proceedings of the CHI Conference on Human Factors in Computing Systems}}. \bibinfo{pages}{1--20}.
\newblock


\bibitem[Khairuddin and Sas(2019)]%
        {khairuddin2019exploration7}
\bibfield{author}{\bibinfo{person}{Irni~Eliana Khairuddin} {and} \bibinfo{person}{Corina Sas}.} \bibinfo{year}{2019}\natexlab{}.
\newblock \showarticletitle{An Exploration of Bitcoin mining practices: Miners' trust challenges and motivations}. In \bibinfo{booktitle}{\emph{Proceedings of the 2019 CHI conference on human factors in computing systems}}. \bibinfo{pages}{1--13}.
\newblock


\bibitem[Knittel et~al\mbox{.}(2019)]%
        {knittel2019most22}
\bibfield{author}{\bibinfo{person}{Megan Knittel}, \bibinfo{person}{Shelby Pitts}, {and} \bibinfo{person}{Rick Wash}.} \bibinfo{year}{2019}\natexlab{}.
\newblock \showarticletitle{" The Most Trustworthy Coin" How Ideological Tensions Drive Trust in Bitcoin}.
\newblock \bibinfo{journal}{\emph{Proceedings of the ACM on Human-Computer Interaction}} \bibinfo{volume}{3}, \bibinfo{number}{CSCW} (\bibinfo{year}{2019}), \bibinfo{pages}{1--23}.
\newblock


\bibitem[Knittel and Wash(2019)]%
        {knittel2019true16}
\bibfield{author}{\bibinfo{person}{Megan~L Knittel} {and} \bibinfo{person}{Rick Wash}.} \bibinfo{year}{2019}\natexlab{}.
\newblock \showarticletitle{How" True Bitcoiners" work on reddit to maintain bitcoin}. In \bibinfo{booktitle}{\emph{Extended Abstracts of the 2019 CHI Conference on Human Factors in Computing Systems}}. \bibinfo{pages}{1--6}.
\newblock


\bibitem[Krause(2025)]%
        {krause2025algorithmic}
\bibfield{author}{\bibinfo{person}{David Krause}.} \bibinfo{year}{2025}\natexlab{}.
\newblock \showarticletitle{Algorithmic Stablecoins: Mechanisms, Risks, and Lessons from the Fall of TerraUSD}.
\newblock \bibinfo{journal}{\emph{Risks, and Lessons from the Fall of TerraUSD (January 11, 2025)}} (\bibinfo{year}{2025}).
\newblock


\bibitem[Kristiansen(2007)]%
        {kristiansen2007rules}
\bibfield{author}{\bibinfo{person}{Kristian Kristiansen}.} \bibinfo{year}{2007}\natexlab{}.
\newblock \showarticletitle{The rules of the game: decentralised complexity and power structures}.
\newblock \bibinfo{journal}{\emph{Socialising Complexity. Structure Interaction and Power in Archaeological Discourse}} (\bibinfo{year}{2007}), \bibinfo{pages}{60--75}.
\newblock


\bibitem[Kusmierz and Overko(2022)]%
        {kusmierz2022centralized}
\bibfield{author}{\bibinfo{person}{Bartosz Kusmierz} {and} \bibinfo{person}{Roman Overko}.} \bibinfo{year}{2022}\natexlab{}.
\newblock \showarticletitle{How centralized is decentralized? Comparison of wealth distribution in coins and tokens}. In \bibinfo{booktitle}{\emph{2022 IEEE International Conference on Omni-layer Intelligent Systems (COINS)}}. IEEE, \bibinfo{pages}{1--6}.
\newblock


\bibitem[law(2022)]%
        {eu22858}
\bibfield{author}{\bibinfo{person}{European~Union law}.} \bibinfo{year}{2022}\natexlab{}.
\newblock \bibinfo{title}{REGULATION (EU) 2022/858 OF THE EUROPEAN PARLIAMENT AND OF THE COUNCIL 2022}.
\newblock
\newblock
\urldef\tempurl%
\url{http://data.europa.eu/eli/reg/2022/858/oj}
\showURL{%
\tempurl}


\bibitem[Long et~al\mbox{.}(2024)]%
        {long2024whales}
\bibfield{author}{\bibinfo{person}{Suwan~Cheng Long}, \bibinfo{person}{Ying Xie}, \bibinfo{person}{Zhengyuan Zhou}, \bibinfo{person}{Brian~M Lucey}, {and} \bibinfo{person}{Andrew Urquhart}.} \bibinfo{year}{2024}\natexlab{}.
\newblock \showarticletitle{From Whales to Waves: The Role of Social Media Sentiment in Shaping Cryptocurrency Markets}.
\newblock \bibinfo{journal}{\emph{Ying and Zhou, Zhengyuan and Lucey, Brian M. and Urquhart, Andrew, From Whales to Waves: The Role of Social Media Sentiment in Shaping Cryptocurrency Markets (January 25, 2024)}} (\bibinfo{year}{2024}).
\newblock


\bibitem[Lustig(2019)]%
        {lustig2019intersecting20}
\bibfield{author}{\bibinfo{person}{Caitlin Lustig}.} \bibinfo{year}{2019}\natexlab{}.
\newblock \showarticletitle{Intersecting imaginaries: visions of decentralized autonomous systems}.
\newblock \bibinfo{journal}{\emph{Proceedings of the ACM on Human-Computer Interaction}} \bibinfo{volume}{3}, \bibinfo{number}{CSCW} (\bibinfo{year}{2019}), \bibinfo{pages}{1--27}.
\newblock


\bibitem[Lyke et~al\mbox{.}(2023)]%
        {lyke2023exploring8}
\bibfield{author}{\bibinfo{person}{Nash Lyke}, \bibinfo{person}{Benjamin~M Gorman}, {and} \bibinfo{person}{Garreth~W Tigwell}.} \bibinfo{year}{2023}\natexlab{}.
\newblock \showarticletitle{Exploring the Accessibility of Crypto Technologies}. In \bibinfo{booktitle}{\emph{Extended Abstracts of the 2023 CHI Conference on Human Factors in Computing Systems}}. \bibinfo{pages}{1--10}.
\newblock


\bibitem[Mai et~al\mbox{.}(2020)]%
        {mai2020user18}
\bibfield{author}{\bibinfo{person}{Alexandra Mai}, \bibinfo{person}{Katharina Pfeffer}, \bibinfo{person}{Matthias Gusenbauer}, \bibinfo{person}{Edgar Weippl}, {and} \bibinfo{person}{Katharina Krombholz}.} \bibinfo{year}{2020}\natexlab{}.
\newblock \showarticletitle{User mental models of cryptocurrency systems-a grounded theory approach}. In \bibinfo{booktitle}{\emph{Sixteenth symposium on usable privacy and security (SOUPS 2020)}}. \bibinfo{pages}{341--358}.
\newblock


\bibitem[Nakamoto(2008)]%
        {nakamoto2008bitcoin}
\bibfield{author}{\bibinfo{person}{Satoshi Nakamoto}.} \bibinfo{year}{2008}\natexlab{}.
\newblock \showarticletitle{Bitcoin: A peer-to-peer electronic cash system}.
\newblock \bibinfo{journal}{\emph{Satoshi Nakamoto}} (\bibinfo{year}{2008}).
\newblock


\bibitem[Nelson(2024)]%
        {nelson2024bitcoin}
\bibfield{author}{\bibinfo{person}{Rob Nelson}.} \bibinfo{year}{2024}\natexlab{}.
\newblock \bibinfo{title}{Can Bitcoin bridge the financial divide for marginalized populations?}
\newblock
\newblock
\urldef\tempurl%
\url{https://www.thestreet.com/crypto/policy/can-bitcoin-bridge-the-financial-divide-for-marginalized-populations}
\showURL{%
\tempurl}
\newblock
\shownote{Accessed: 2025-01-22}.


\bibitem[Preukschat and Reed(2018)]%
        {preukschat2018blockchain}
\bibfield{author}{\bibinfo{person}{Alex Preukschat} {and} \bibinfo{person}{Daniel Reed}.} \bibinfo{year}{2018}\natexlab{}.
\newblock \bibinfo{booktitle}{\emph{Blockchain: Distributed Ledger Technology and its Implications}}.
\newblock \bibinfo{publisher}{Manning Publications}.
\newblock


\bibitem[Prewett et~al\mbox{.}(2020)]%
        {prewett2020blockchain}
\bibfield{author}{\bibinfo{person}{Kyleen~W Prewett}, \bibinfo{person}{Gregory~L Prescott}, {and} \bibinfo{person}{Kirk Phillips}.} \bibinfo{year}{2020}\natexlab{}.
\newblock \showarticletitle{Blockchain adoption is inevitable—Barriers and risks remain}.
\newblock \bibinfo{journal}{\emph{Journal of Corporate accounting \& finance}} \bibinfo{volume}{31}, \bibinfo{number}{2} (\bibinfo{year}{2020}), \bibinfo{pages}{21--28}.
\newblock


\bibitem[Romiti et~al\mbox{.}(2019)]%
        {romiti2019deep}
\bibfield{author}{\bibinfo{person}{Matteo Romiti}, \bibinfo{person}{Aljosha Judmayer}, \bibinfo{person}{Alexei Zamyatin}, {and} \bibinfo{person}{Bernhard Haslhofer}.} \bibinfo{year}{2019}\natexlab{}.
\newblock \showarticletitle{A deep dive into bitcoin mining pools: An empirical analysis of mining shares}.
\newblock \bibinfo{journal}{\emph{arXiv preprint arXiv:1905.05999}} (\bibinfo{year}{2019}).
\newblock


\bibitem[Saldivar et~al\mbox{.}(2023)]%
        {saldivar2023blockchain1}
\bibfield{author}{\bibinfo{person}{Jorge Saldivar}, \bibinfo{person}{Elena Mart{\'\i}nez-Vicente}, \bibinfo{person}{David Rozas}, \bibinfo{person}{Maria-Cruz Valiente}, {and} \bibinfo{person}{Samer Hassan}.} \bibinfo{year}{2023}\natexlab{}.
\newblock \showarticletitle{Blockchain (not) for everyone: Design challenges of blockchain-based applications}. In \bibinfo{booktitle}{\emph{Extended Abstracts of the 2023 CHI Conference on Human Factors in Computing Systems}}. \bibinfo{pages}{1--8}.
\newblock


\bibitem[Sas and Khairuddin(2017)]%
        {sas2017design9}
\bibfield{author}{\bibinfo{person}{Corina Sas} {and} \bibinfo{person}{Irni~Eliana Khairuddin}.} \bibinfo{year}{2017}\natexlab{}.
\newblock \showarticletitle{Design for trust: An exploration of the challenges and opportunities of bitcoin users}. In \bibinfo{booktitle}{\emph{Proceedings of the 2017 CHI Conference on Human Factors in Computing Systems}}. \bibinfo{pages}{6499--6510}.
\newblock


\bibitem[Scharfman(2023)]%
        {scharfman2023cryptocurrency}
\bibfield{author}{\bibinfo{person}{Jason Scharfman}.} \bibinfo{year}{2023}\natexlab{}.
\newblock \bibinfo{booktitle}{\emph{The Cryptocurrency and Digital Asset Fraud Casebook}}.
\newblock \bibinfo{publisher}{Springer}.
\newblock


\bibitem[Schilling and Uhlig(2019)]%
        {schilling2019decentralized}
\bibfield{author}{\bibinfo{person}{Linda Schilling} {and} \bibinfo{person}{Harald Uhlig}.} \bibinfo{year}{2019}\natexlab{}.
\newblock \showarticletitle{Decentralized Finance: Opportunities and Challenges}.
\newblock \bibinfo{journal}{\emph{Journal of Financial Technology}} \bibinfo{volume}{5}, \bibinfo{number}{1} (\bibinfo{year}{2019}), \bibinfo{pages}{23--45}.
\newblock


\bibitem[Schmid and Shestakov(2024)]%
        {schmid2024blockchain}
\bibfield{author}{\bibinfo{person}{Stefan Schmid} {and} \bibinfo{person}{Dmitry Shestakov}.} \bibinfo{year}{2024}\natexlab{}.
\newblock \showarticletitle{Blockchain Governance and Liquid Democracy--Quantifying Decentralization in Gitcoin and Internet Computer}. In \bibinfo{booktitle}{\emph{Proceedings of the 2024 Workshop on Advanced Tools, Programming Languages, and PLatforms for Implementing and Evaluating algorithms for Distributed systems}}. \bibinfo{pages}{1--7}.
\newblock


\bibitem[Si et~al\mbox{.}(2024)]%
        {si2024understanding11}
\bibfield{author}{\bibinfo{person}{Janice~Jianing Si}, \bibinfo{person}{Tanusree Sharma}, {and} \bibinfo{person}{Kanye~Ye Wang}.} \bibinfo{year}{2024}\natexlab{}.
\newblock \showarticletitle{Understanding User-Perceived Security Risks and Mitigation Strategies in the Web3 Ecosystem}. In \bibinfo{booktitle}{\emph{Proceedings of the CHI Conference on Human Factors in Computing Systems}}. \bibinfo{pages}{1--22}.
\newblock


\bibitem[Sristy(2021)]%
        {sristy2021blockchain}
\bibfield{author}{\bibinfo{person}{Archana Sristy}.} \bibinfo{year}{2021}\natexlab{}.
\newblock \bibinfo{title}{Blockchain in the food supply chain - What does the future look like?}
\newblock
\newblock
\urldef\tempurl%
\url{https://tech.walmart.com/content/walmart-global-tech/en_us/blog/post/blockchain-in-the-food-supply-chain.html}
\showURL{%
\tempurl}
\newblock
\shownote{Walmart Global Tech Blog}.


\bibitem[Swan(2015)]%
        {swan2015blockchain}
\bibfield{author}{\bibinfo{person}{Melanie Swan}.} \bibinfo{year}{2015}\natexlab{}.
\newblock \bibinfo{booktitle}{\emph{Blockchain: Blueprint for a New Economy}}.
\newblock \bibinfo{publisher}{O'Reilly Media, Inc.}
\newblock


\bibitem[Tapscott and Tapscott(2016)]%
        {tapscott2016blockchain}
\bibfield{author}{\bibinfo{person}{Don Tapscott} {and} \bibinfo{person}{Alex Tapscott}.} \bibinfo{year}{2016}\natexlab{}.
\newblock \bibinfo{booktitle}{\emph{Blockchain Revolution: How the Technology Behind Bitcoin Is Changing Money, Business, and the World}}.
\newblock \bibinfo{publisher}{Portfolio}.
\newblock


\bibitem[Tyma et~al\mbox{.}(2022)]%
        {tyma2022understanding}
\bibfield{author}{\bibinfo{person}{Bridget Tyma}, \bibinfo{person}{Rina Dhillon}, \bibinfo{person}{Prabhu Sivabalan}, {and} \bibinfo{person}{Bernhard Wieder}.} \bibinfo{year}{2022}\natexlab{}.
\newblock \showarticletitle{Understanding accountability in blockchain systems}.
\newblock \bibinfo{journal}{\emph{Accounting, Auditing \& Accountability Journal}} \bibinfo{volume}{35}, \bibinfo{number}{7} (\bibinfo{year}{2022}), \bibinfo{pages}{1625--1655}.
\newblock


\bibitem[{UN Innovation Network}(2024)]%
        {uninnovation2024blockchain}
\bibfield{author}{\bibinfo{person}{{UN Innovation Network}}.} \bibinfo{year}{2024}\natexlab{}.
\newblock \bibinfo{title}{Blockchain and Distributed Ledger Technologies}.
\newblock
\newblock
\urldef\tempurl%
\url{https://www.uninnovation.network/innovation-areas/blockchain-distributed-ledger-technologies}
\showURL{%
\tempurl}
\newblock
\shownote{Accessed: 2024-01-21}.


\bibitem[Vie(2008)]%
        {vie2008digital}
\bibfield{author}{\bibinfo{person}{Stephanie Vie}.} \bibinfo{year}{2008}\natexlab{}.
\newblock \showarticletitle{Digital divide 2.0:“Generation M” and online social networking sites in the composition classroom}.
\newblock \bibinfo{journal}{\emph{Computers and Composition}} \bibinfo{volume}{25}, \bibinfo{number}{1} (\bibinfo{year}{2008}), \bibinfo{pages}{9--23}.
\newblock


\bibitem[Voskobojnikov et~al\mbox{.}(2021a)]%
        {voskobojnikov2021balancing}
\bibfield{author}{\bibinfo{person}{Artemij Voskobojnikov}, \bibinfo{person}{Volker Skwarek}, \bibinfo{person}{Atefeh Mashatan}, \bibinfo{person}{Shin’Ichiro Matsuo}, \bibinfo{person}{Chris Rowell}, {and} \bibinfo{person}{Tim Weing{\"a}rtner}.} \bibinfo{year}{2021}\natexlab{a}.
\newblock \showarticletitle{Balancing Security: A Moving Target}.
\newblock \bibinfo{journal}{\emph{Building Decentralized Trust: Multidisciplinary Perspectives on the Design of Blockchains and Distributed Ledgers}} (\bibinfo{year}{2021}), \bibinfo{pages}{63--94}.
\newblock


\bibitem[Voskobojnikov et~al\mbox{.}(2021b)]%
        {voskobojnikov2021u12}
\bibfield{author}{\bibinfo{person}{Artemij Voskobojnikov}, \bibinfo{person}{Oliver Wiese}, \bibinfo{person}{Masoud Mehrabi~Koushki}, \bibinfo{person}{Volker Roth}, {and} \bibinfo{person}{Konstantin Beznosov}.} \bibinfo{year}{2021}\natexlab{b}.
\newblock \showarticletitle{The u in crypto stands for usable: An empirical study of user experience with mobile cryptocurrency wallets}. In \bibinfo{booktitle}{\emph{Proceedings of the 2021 CHI Conference on Human Factors in Computing Systems}}. \bibinfo{pages}{1--14}.
\newblock


\bibitem[{Web3devs}(2024)]%
        {web3devs2024}
\bibfield{author}{\bibinfo{person}{{Web3devs}}.} \bibinfo{year}{2024}\natexlab{}.
\newblock \bibinfo{title}{Cryptocurrency Market Analysis for 2024: Key Trends and Insights}.
\newblock
\newblock
\urldef\tempurl%
\url{https://web3devs.com/blog/cryptocurrency-market-analysis-2024/}
\showURL{%
\tempurl}
\newblock
\shownote{Accessed: 2025-01-22}.


\bibitem[Werbach(2018)]%
        {werbach2018blockchain}
\bibfield{author}{\bibinfo{person}{Kevin Werbach}.} \bibinfo{year}{2018}\natexlab{}.
\newblock \showarticletitle{The Blockchain and the New Architecture of Trust}.
\newblock \bibinfo{journal}{\emph{MIT Technology Review}} \bibinfo{volume}{121}, \bibinfo{number}{3} (\bibinfo{year}{2018}).
\newblock


\bibitem[Xiao et~al\mbox{.}(2024)]%
        {xiao2024centralized21}
\bibfield{author}{\bibinfo{person}{Yunpeng Xiao}, \bibinfo{person}{Bufan Deng}, \bibinfo{person}{Siqi Chen}, \bibinfo{person}{Kyrie~Zhixuan Zhou}, \bibinfo{person}{Ray LC}, \bibinfo{person}{Luyao Zhang}, {and} \bibinfo{person}{Xin Tong}.} \bibinfo{year}{2024}\natexlab{}.
\newblock \showarticletitle{" Centralized or Decentralized?": Concerns and Value Judgments of Stakeholders in the Non-Fungible Tokens (NFTs) Market}.
\newblock \bibinfo{journal}{\emph{Proceedings of the ACM on Human-Computer Interaction}} \bibinfo{volume}{8}, \bibinfo{number}{CSCW1} (\bibinfo{year}{2024}), \bibinfo{pages}{1--34}.
\newblock


\bibitem[Yu et~al\mbox{.}(2024)]%
        {yu2024don10}
\bibfield{author}{\bibinfo{person}{Yaman Yu}, \bibinfo{person}{Tanusree Sharma}, \bibinfo{person}{Sauvik Das}, {and} \bibinfo{person}{Yang Wang}.} \bibinfo{year}{2024}\natexlab{}.
\newblock \showarticletitle{" Don't put all your eggs in one basket": How Cryptocurrency Users Choose and Secure Their Wallets}. In \bibinfo{booktitle}{\emph{Proceedings of the CHI Conference on Human Factors in Computing Systems}}. \bibinfo{pages}{1--17}.
\newblock


\bibitem[Zhang et~al\mbox{.}(2023)]%
        {zhang2023data}
\bibfield{author}{\bibinfo{person}{Wenxiang Zhang}, \bibinfo{person}{Saeed Siyal}, \bibinfo{person}{Samina Riaz}, \bibinfo{person}{Riaz Ahmad}, \bibinfo{person}{Mohd~Faiz Hilmi}, {and} \bibinfo{person}{Zhi Li}.} \bibinfo{year}{2023}\natexlab{}.
\newblock \showarticletitle{Data security, customer trust and intention for adoption of Fintech services: an empirical analysis from commercial bank users in Pakistan}.
\newblock \bibinfo{journal}{\emph{Sage Open}} \bibinfo{volume}{13}, \bibinfo{number}{3} (\bibinfo{year}{2023}), \bibinfo{pages}{21582440231181388}.
\newblock


\bibitem[Zheng et~al\mbox{.}(2018)]%
        {zheng2018overview}
\bibfield{author}{\bibinfo{person}{Zibin Zheng}, \bibinfo{person}{Shaoan Xie}, \bibinfo{person}{Hong-Ning Dai}, \bibinfo{person}{Xiangping Chen}, {and} \bibinfo{person}{Huainian Wang}.} \bibinfo{year}{2018}\natexlab{}.
\newblock \showarticletitle{An Overview of Blockchain Technology: Architecture, Consensus, and Future Trends}.
\newblock \bibinfo{journal}{\emph{IEEE International Conference on Big Data (Big Data)}} (\bibinfo{year}{2018}), \bibinfo{pages}{557--564}.
\newblock


\end{thebibliography}

\begin{appendix}
\section{Interview Protocol and Recruitment Strategy}

\subsection{Interview Format}
Each interview followed a semi-structured format, enabling flexibility to probe emergent themes while maintaining consistency across participants. The interviews were conducted over video calls and lasted 60–90 minutes. Below are example questions asked:

\begin{itemize}
  \item How did you first get involved with cryptocurrency or\\blockchain technologies?
  \item What platforms or tools do you currently use to manage your crypto assets?
  \item What does decentralization mean to you, and how important is it in your decision-making?
  \item Have you ever had concerns about the security or trustworthiness of platforms you use?
  \item Can you recall a moment when your trust in a platform or technology changed significantly?
  \item How do you balance convenience and control when choosing between centralized and decentralized services?
\end{itemize}

\subsection{Sampling Strategy and Recruitment}
We employed purposive sampling to recruit participants representing diverse roles in the blockchain ecosystem, including traders, investors, developers, researchers, and enthusiasts. Participants were recruited through blockchain forums, social media (e.g., X \textit{(formerly Twitter}), LinkedIn), and professional networks. Snowball sampling was also used, where existing participants referred others with relevant experience. All participants were screened to ensure at least two years of involvement with cryptocurrency or blockchain technologies. Before each interview, participants were provided with an information sheet and consent form, outlining the purpose of the study, confidentiality measures, and their right to withdraw at any point. Participant demographics and platform use are summarized in Table~\ref{tab:participants} (Section 4.1).

\section{Thematic Coding Categories}
Our interview analysis resulted in 28 unique codes grouped under six thematic categories. Table \ref{tab:codes} lists each category along with its associated codes, definitions, and illustrative quotes drawn from participant interviews.

\begin{table*}[]
\centering
\caption{Thematic Categories, Codes, Definitions, and Example Quotes}
\label{tab:codes}
\renewcommand{\arraystretch}{1.4} 
\small 
\resizebox{\textwidth}{!}{%
\begin{tabular}{@{}p{3cm}p{3.5cm}p{5.5cm}p{6.5cm}@{}}
\toprule
\textbf{Category} & \textbf{Code} & \textbf{Definition} & \textbf{Example Quote} \\
\midrule

\multirow{4}{*}{Trust in Decentralization}
& Centralization Concerns & Distrust of custodial or centralized blockchain services & “I still keep most of my assets on Coinbase—hardware wallets scare me.” \\
& Stakeholder Influence & Perceived control by dominant actors or institutions & “If a few big miners can decide everything, what’s the point of decentralization?” \\
& Platform Governance & Opinions about voting rights and on-chain decisions & “I feel like my vote doesn’t matter in most DAOs.” \\
& Transparency Expectations & Trust rooted in visible, verifiable data flows & “The fact that I can audit everything is what gives me confidence.” \\

\midrule

\multirow{4}{*}{Security Practices}
& Key Management & Strategies and anxieties around seed phrase handling & “I’ve written my seed down in three different places.” \\
& Exchange Hacks & Reactions to breaches at centralized platforms & “After Mt. Gox and FTX, I stopped trusting centralized exchanges.” \\
& Two-Factor Authentication & Use of extra authentication layers & “I always enable 2FA, even if it’s annoying.” \\
& Risk Containment & Spreading assets to mitigate potential loss & “I never keep all my coins in one place anymore.” \\

\midrule

\multirow{4}{*}{Platform Choice}
& Usability Preference & Choosing tools for ease of use over decentralization & “Uniswap is great, but Coinbase is just easier to use.” \\
& Speed and Efficiency & Preference for platforms with faster transactions & “If it takes 10 minutes to confirm, I’m out.” \\
& Cost Sensitivity & Concerns about transaction/gas fees & “I avoid Ethereum when gas is too high—it’s just not worth it.” \\
& Familiarity & Comfort from using known or mainstream platforms & “I started with Binance and just stuck with it.” \\

\midrule

\multirow{4}{*}{Ideological Commitment}
& Philosophical Alignment & Expressions of belief in decentralization ideals & “Decentralization is about freedom from banks.” \\
& Pragmatic Tradeoffs & Accepting centralization for convenience or safety & “I know it’s not pure, but it works.” \\
& Anti-Establishment Sentiment & Distrust of governments or banks & “The whole point is to not rely on corrupt institutions.” \\
& Crypto Ethos & Reference to values like censorship-resistance or openness & “If it’s not open-source, I don’t trust it.” \\

\midrule

\multirow{4}{*}{Learning and Onboarding}
& Self-Education & Learning through YouTube, blogs, or forums & “Most of what I know came from Reddit and YouTube.” \\
& Trial and Error & Learning by making mistakes & “I lost \$50 once because I copied the wrong address.” \\
& Community Support & Relying on peer knowledge & “I usually ask in Discord if I’m unsure.” \\
& Overload and Confusion & Feeling overwhelmed by complex concepts & “There’s so much jargon—it took me months to understand wallets.” \\

\midrule

\multirow{4}{*}{Trust Signals}
& Reputation Systems & Using reviews or trust scores & “I always check Reddit before trying a new platform.” \\
& Audits and Certifications & Valuing third-party security verification & “If it’s unaudited, I’m not touching it.” \\
& Branding and UX & Judging legitimacy based on interface polish & “If the UI looks shady, I don’t even register.” \\
& Social Proof & Following where others go & “If enough people I trust use it, I’ll try it too.” \\

\bottomrule
\end{tabular}}
\end{table*}

\end{appendix}

\end{document}